\documentclass[journal]{IEEEtran}

\usepackage{ifpdf}
\usepackage{graphicx}
\usepackage{amssymb}
\usepackage[tbtags]{amsmath}
\usepackage{amsmath}

\usepackage{cite}
\usepackage{epstopdf}
\usepackage{dsfont}
\usepackage[justification=centering]{caption}
\usepackage{color}
\usepackage{algorithm}
\usepackage{algorithmicx, algpseudocode}
\usepackage{etoolbox}
\usepackage{subcaption}
\usepackage{enumitem}

\usepackage{balance}

\newtheorem{theorem}{Theorem}

\newtheorem{corollary}{Corollary}

\makeatletter
\def\ScaleIfNeeded{%
\ifdim\Gin@nat@width>\linewidth \linewidth \else \Gin@nat@width \fi
} \makeatother

\newcommand{\mbf}[1]{\mathbf{#1}}
\begin{document}

\title{Joint Wireless and Computing Resources Allocation in Multi-Cell MEC}




\author{\IEEEauthorblockN{Ming Zeng} \\
\IEEEauthorblockA{Department of Electrical and Computer Engineering, Laval University, Quebec, Canada }

\IEEEauthorblockA {Email: ming.zeng@gel.ulaval.ca}
}

\maketitle

\begin{abstract}
This paper addresses join wireless and computing resource allocation in mobile edge computing (MEC) systems with several access points and with the possibility that users connect to many access points, and utilize the computation capability of many servers at the same time. The problem of sum transmission energy minimization under response time constraints is considered. It is proved, that the optimization problem is non-convex. The complexity of optimization of a part of the system parameters is investigated, and based on these results an Iterative Resource Allocation procedure is proposed, that converges to a local optimum. The performance of the joint resource allocation is evaluated by comparing it to lower and upper bounds defined by less or more flexible multi-cell MEC architectures. The results show  that the free selection of the access point is crucial for good system performance. 
\end{abstract}

\IEEEpeerreviewmaketitle

\section{Introduction}
Mobile edge computing (MEC) represents a promising technology to reduce the latency for 5G and beyond networks \cite{ Y_Mao, Zeng_VTM20}. It allows users to offload their computational intensive task to servers in close proximity, and thus, significantly enhancing their computation capacities and prolonging their lifespan. The whole process in general consists of the following three sequential phases \cite{ A_WCL5}: 1) uplink data transmission; 2) task processing at the MEC server; and 3) results feedback through the downlink. The uplink and downlink transmission phases involve the allocation of wireless resources, while the task processing phase concerns the computational resources. As a result, a joint allocation of wireless and computational resources is required to deliver a decent system performance\cite{BarbarossaSPM14}.

Initially, MEC was applied to single-user systems, where the optimization variables include
the transmission power, the offloading decision and ratio as well as the central processing unit (CPU)
frequency for computing \cite{W_zhang_TWC13, Y_wang_COM16}.
The single-user systems were then extended to the multi-user ones \cite{Xchen, Cyou, Ming_PIMRC}, where how to share the wireless and computational resources among the users directly affect the system performance. 
Recently, non-orthogonal multiple access is envisioned as a promising access technology \cite{Hao_TCOM19}, and its application to MEC has been considered, for example in \cite{Zeng_Adhoc20}. 
Note that the above mentioned works only apply to single-cell systems. The general scenario of multi-cell systems is attracting great attention recently \cite{Proa_INF19, Yang_Access19, Zeng_WCL2019}.
Compared with single-cell systems, a new problem that emerges in multi-cell systems is how to match the users to the access points (APs).

In this paper we consider a multi-AP multi-user MEC system, where each user can access multiple APs and utilize their computation resource. This generalizes the previous works \cite{Proa_INF19, Yang_Access19, Zeng_WCL2019}, where each user only offloads to one AP. Meanwhile, as in \cite{Zeng_WCL2019}, we consider the general case with flexible bandwidth allocation across both the APs and users. The system objective is to minimize the sum energy consumption under response time constraints. The formulated problem is shown to be non-convex. To handle it, we first investigate the complexity of optimization of a part of the system parameters. On this basis, we propose an iterative resource allocation procedure that converges to local optimum. To evaluate the proposed iterative solution, we compare it with the lower and upper bounds defined by less or more flexible multi-cell MEC architectures. Presented results validate the necessity of free selection of APs. Meanwhile, binary allocation, where all users select the AP with the highest share of their load provides a performance close to parallel processing. This facilitates its application to large systems, where the level of parallel processing is low.

\section{System Model}
\label{systemmodel}
We consider a multi-cell MEC system that consists of $K$ users, and $M$ APs, each equipped with a MEC server. We denote the set of users by $\mathcal{K} = \{1, \cdots, K\}$, and APs by $\mathcal{M} = \{1, \cdots, M\}$. We consider that each user $i \in \mathcal{K}$ generates a computationally intensive {\color{black} and delay sensitive} task, which is characterized by {\color{black} three} parameters, the size $L_i$ of the input data, the number $W_i$ of CPU cycles required to perform the computation, and the completion time constraint $D_i$.
To save energy consumption at the user for processing the tasks, and satisfy the delay constraint, 
{\color{black}each user offloads its computing task to one or multiple APs for processing. That is, each user $i \in \mathcal{K}$ offloads part of its input data, i.e., $L_{i,j}$ to AP $j, j \in \mathcal{M}$, satisfying $L_{i,j}\geq 0$ and $\sum_{j=1}^M L_{i,j}=L_i$.} 
For simplicity we assume that this parallel processing has no communication or computation overhead \cite{Ref_27}. 
The objective of the considered MEC system is to minimize the energy consumption for data transmission under the delay constraint, by jointly allocating the data to be sent to the APs, the wireless and the computing resources.




\subsection{Wireless resource management}
The overall system bandwidth is $B$ Hz, which should be appropriately shared among the users. We consider flat fading channel and orthogonal access with frequency division multiple access. 
Denote the corresponding channel gain for user $i$ to AP $j$ by $h_{i,j}$. Then, the achievable data rate at user $i$ to AP $j$ is given by {\color{black}
\begin{equation} \label{rate}
R_{i,j}=x_{i,j} \log_2 \left( 1+\frac{P_{i,j} h_{i,j}}{x_{i,j} N_0} \right),
\end{equation}
where $P_{i,j}$ is the corresponding transmission power, while $x_{i,j}$ denotes the allocated bandwidth, satisfying $\sum_{i \in \mathcal{K}} \sum_{j \in \mathcal{M}}  x_{i,j} =B$.
Besides, $N_0$ is the noise power spectral density coefficient.

Accordingly, the transmission time and the resulting transmission energy consumption are respectively given by
\begin{equation} \label{T_E}
T_{i,j}=\frac{L_{i,j}}{R_{i,j}}~ {\rm{and}}~ E_{i,j}=\frac{L_{i,j} P_{i,j}}{R_{i,j}}.
\end{equation}

\subsection{Computing resource management}
Let us denote the computational capacity of the MEC server for AP $j, j\in \mathcal{M}$ by $C_j$.
We denote the computing resource allocated to user $i$ as $q_{i,j}$, satisfying $\sum_{i \in \mathcal{K}} q_{i,j} = C_j$.
Then, the computational time of user $i$'s task is given by $Q_{i,j}=\frac{W_{i,j}}{q_{i,j}}$. We assume that there is a linear relation between $L_{i,j}$ and $W_{i,j}$, i.e., $W_{i,j}=\eta L_{i,j}$, where $\eta$ is the coefficient \cite{Ref_4}. Then, we have 
\begin{equation} \label{Q_ij}
Q_{i,j}=\frac{\eta L_{i,j}}{q_{i,j}}.
\end{equation}


\section{Problem Formulation and General Results}

We consider the problem of total transmission energy minimization, under the constraint on the completion time of the computational tasks. That is, for each user $i$, the sum of the transmission and computational times should not
violate the maximum delay $D_i$, i.e., $T_{i,j}+Q_{i,j} \leq D_i, \forall j \in \mathcal{M}$.
{\color{black}We disregard the time needed for the downlink transmission of the results, since it concerns usually small amounts of data
\cite{L_IoT11, Latency, F_TWC9, Zeng_WoWMoM19}.
Additionally, we do not consider the energy consumption of the computation at the MEC servers, since it is independent from the resource allocation (i.e., all computing needs to be performed at the MEC servers anyway, and consumes the same energy \cite{ F_TWC9}).
}

The delay constraint then can be turned into the following rate requirement:
\begin{equation}
R_{i,j} \geq \frac{L_{i,j}}{D_i - Q_{i,j}}, \forall i \in \mathcal{K}, j \in \mathcal{M}.
\end{equation}

The resource allocation concerns the allocation of bandwidth, power, the computing resource and the data for each user on each AP. The energy minimization problem can be formulated as
\begin{subequations} \label{P1}
\begin{align}
\text{P1}:& ~ \underset{\mbf{P},\mbf{x},\mbf{q}, \mbf{L}}{\text{min}} \sum_{i \in \mathcal{K}} \sum_{j \in \mathcal{M}} E_{i,j} \\
\text{s.t.}~~ & ~R_{i,j} \geq \frac{L_{i,j}}{D_i - Q_{i,j}}, \forall i \in \mathcal{K}, j \in \mathcal{M} \\
& ~\sum_{i \in \mathcal{K}} \sum_{j \in \mathcal{M}}  x_{i,j} =B  \\
& ~ \sum_{i \in \mathcal{K}} q_{i,j} = C_j, \forall j \in \mathcal{M} \\
&~ \sum_{j=1}^M L_{i,j}=L_i, \forall i \in \mathcal{K}
\end{align}
\end{subequations}
where $\mbf{P} \in \mathbb{R}^{K},\mbf{x} \in \mathbb{R}^{K},\mbf{q} \in \mathbb{R}^{K}, \mbf{L} \in \mathbb{R}^{K \times M}$ are the matrix of allocated powers $P_{i,j}$, bandwidth $x_{i,j}$, computational resource $q_{i,j}$ and data size $L_{i,j}$, respectively.
Inequality constraints (\ref{P1}b) reflect the minimum data rate requirement for each user on each AP. Constraints (\ref{P1}c) limit the bandwidth, while (\ref{P1}d) restrict the computing resource. Constraints (\ref{P1}e) limit the data size.




To solve P1, we need to jointly allocate the wireless and computing resources and data, which are coupled in a non-linear way through the delay constraint. To progress, we first state the following theorem.

\begin{theorem}\label{theorem_power}
Under any given bandwidth, computing resource and data size allocation $\mbf{x},\mbf{q}$, $\mbf{L}$, the energy consumption is minimized when $T_{i,j}+Q_{i,j}=D_i$, $\forall i \in \mathcal{K}, j \in \mathcal{M}$ holds and the transmission power is set as
\begin{equation}\label{eq:pum_OMA}
P_{i,j}=\frac{ N_0  x_{i,j}}{h_{i,j}} { \left(2^{\frac{R_{i,j}^{\min}}{x_{i,j}}}-1 \right)},
\forall i \in \mathcal{K}, j \in \mathcal{M}
\end{equation}
where $R_{i,j}^{\rm{min}}$ is the minimum rate that still fulfills the delay requirement, i.e., $R_{i,j}^{\rm{min}}=\frac{L_{i,j}}{D_i-Q_{i,j}}$.
\end{theorem}

\begin{IEEEproof}
When $x_{i,j}$, $q_{i,j}$ and $L_{i,j}$ are given, the energy consumption of the users is independent, and minimizing the total energy consumption is equivalent to minimizing that of each user on each AP. Without loss of generality, we look at $E_{i,j}$, which can be reformulated as
\begin{equation} \label{power_consumption}
 E_{i,j}=  \frac{L_{i,j} P_{i,j} }{R_{i,j}}=\frac{L_{i,j} P_{i,j} }{x_{i,j} \log_2 \left( 1+ \frac{P_{i,j} h_{i,j}}{x_{i,j} N_0} \right)}.
\end{equation}

According to \eqref{power_consumption}, it is clear that $E_{i,j}$ increases with $P_{i,j}$. Therefore, $E_{i,j}$ is minimized when the minimum power is used. Meanwhile, to satisfy the delay constraint, we have $R_{i,j}= x_{i,j} \log_2 \left( 1+ \frac{P_{i,j} h_{i,j}}{x_{i,j} N_0} \right) \geq R_{i,j}^{\rm{min}}$, i.e., $P_{i,j} \geq {(2^{R_{i,j}^{\rm{min}}/x_{i,j}} -1)  N_0 x_{i,j}}/{ h_{i,j}}$. 
At equality the achieved rate is $R_{i,j}^{\rm{min}}$, which in turn results a transmission time of $T_{i,j}=D_i-Q_{i,j}$.
This concludes the proof.
\end{IEEEproof}

Based on Theorem \ref{theorem_power}, Problem
P1 can be simplified as
\begin{subequations} \label{P2}
\begin{align}
\text{P2}:& ~ \underset{\mbf{x},\mbf{q}, \mbf{L}}{\text{min}} \sum_{i \in \mathcal{K}} \sum_{j \in \mathcal{M}} \frac{ N_0  x_{i,j}}{h_{i,j}} \left(D_{i}-\frac{\eta L_{i,j}}{q_{i,j}} \right) \left(2^{\frac{L_{i,j}/x_{i,j}} {D_{i}-\frac{\eta L_{i,j}}{q_{i,j}}}}-1 \right)\\
\text{s.t.}~~
& ~\sum_{i \in \mathcal{K}} \sum_{j \in \mathcal{M}}  x_{i,j} =B  \\
& ~ \sum_{i \in \mathcal{K}} q_{i,j} = C_j, \forall j \in \mathcal{M} \\
&~ \sum_{j=1}^M L_{i,j}=L_i, \forall i \in \mathcal{K}
\end{align}
\end{subequations}
This is the problem that we will evaluate in detail.

\begin{theorem}\label{theorem_main_conv}
Problem P2, that is, the optimal joint allocation of bandwidth $\mbf{x}$, computing resource $\mbf{q}$ and data size $\mbf{L}$ is a non-convex problem.
\end{theorem}
\begin{IEEEproof}
P2 requires to jointly allocate bandwidth, computing resource and data size, and is difficult to handle.
In the following, we show that P2 is actually non-convex. Without loss of generality, we only consider $E_{i,j}$. We first consider $E_{i,j}$ over $x_{i,j}$ and $L_{i,j}$.
Denote its Hessian matrix as $\mbf{H}_{i,j}$, which is \begin{align}
\mbf{H}_{i,j} = &
  \frac{N_0}{h_i} 2 ^{\frac{L_{i,j}/x_{i,j}}{D_{i}-\eta L_{i,j} /q_{i,j}}} \ln2^2 \\ \nonumber
  & \times
   \begin{bmatrix}
    \frac{L_{i,j}}{(D_{i}-\eta L_{i,j} /q_{i,j}) x_{i,j}^3}
     & - \frac{D_{i} L_{i,j}}{(D_{i}-\eta L_{i,j} /q_{i,j})^3 x_{i,j}^2} \\
    - \frac{D_{i} L_{i,j}}{(D_{i}-\eta L_{i,j} /q_{i,j})^3 x_{i,j}^2}
     & \frac{D_{i}^2}{(D_{i}-\eta L_{i,j} /q_{i,j})^3 x_{i,j}}
  \end{bmatrix}.
\end{align}

It can be verified that $\rm{det}(\mbf{H}_{i,j}) \geq 0$ does not always hold. {\color{black}For example,
by setting $ D_i=0.1, \eta=0.5, L_{i,j}=0.1, x_{i,j}=0.1, q_{i,j}=1, \frac{N_0}{h_i}=1 $, we obtain $\rm{det}(\mbf{H}_{i,j})=-3.1437 \times 10^{13}<0 $.}
Since all leading principle minors of a convex function should be greater or equal than $0$, this indicates P2 is non-convex.
\end{IEEEproof}

{\color{black}
\section{Complexity of Subproblems and the Joint Resource Allocation Algorithm}

Problem P2 considers three decision variables, $\mbf{L},$ $\mbf{q}$ and $\mbf{x}$. In this section we provide a discussion about the complexity of the subproblems, where some of the variables are considered as given input parameters. This discussion helps us to find ways for decomposing the optimization problem into tractable subproblems. It also guides future system design, where the network, the computing resources and the application may be controlled by three different parties.

\subsection{Complexity of Subproblems}

}

Let us first consider subproblems of P2 with a single free decision variable.

\begin{theorem}\label{theorem-singleparam}
Subproblems of P2, where two of the three variables of $\mbf{L},$ $\mbf{q}$ and $\mbf{x}$ are fixed, are convex problems.
\end{theorem}
\begin{IEEEproof}
To prove the theorem, we need to consider the following three subcases:
\begin{enumerate}[label=\alph*)]
\item Optimizing data size $\mbf{L}$. According to $\mbf{H}_{i,j}$, we have $$\frac{\partial^2 E_{i,j}}{\partial L_{i,j}^2}= \frac{N_0 D_{i}^2}{h_{i,j}x_{i,j}(D_{i}-\eta L_{i,j} /q_{i,j})^3} 2 ^{\frac{L_{i,j}/x_{i,j}}{D_{i}-\eta L_{i,j} /q_{i,j}}} \ln2^2 >0.$$ Therefore, the considered problem is convex.
\item Optimizing bandwidth $\mbf{x}$. According to $\mbf{H}_{i,j}$, we have
$$\frac{\partial^2 E_{i,j}}{\partial x_{i,j}^2}=   \frac{L_{i,j}}{h_i(D_{i}-\eta L_{i,j} /q_{i,j}) x_{i,j}^3} 2 ^{\frac{N_0 L_{i,j}/x_{i,j}}{ D_{i}-\eta L_{i,j} /q_{i,j}}} \ln2^2>0. $$ Therefore, the considered problem is also convex.
\item Optimizing computing resource $\mbf{q}$. We replace variables $q_{i,j}$ with $t_{i,j}=D_i-{\eta L_{i,j}}/{q_{i,j}} $ to get
\begin{subequations} \label{Pq}
\begin{align}
 \underset{ {\mbf{t}}}{\text{min}} & ~ \sum_{i \in \mathcal{K}} \frac{  N_0 }{h_{i,j}} x_{i,j} t_{i,j} \left(2^{\frac{L_{i,j}}{x_{i,j} t_{i,j}}}-1 \right) \\
\text{s.t.}
& ~ \sum_{i \in \mathcal{K} } \frac{\eta L_{i,j}}{D_i-t_{i,j}} = C_j, \forall j \in \mathcal{M}
\end{align}
\end{subequations}

In \eqref{Pq}, equality (\ref{Pq}b) is clearly not affine, and thus, the feasible set is non-convex.
To address it, we relax the equality constraint and substitute (\ref{Pq}b) with
\begin{equation} \label{modified C2_new_q}
\sum_{i \in \mathcal{K}} \frac{\eta L_{i,j}}{D_i-t_{i,j}} \leq C_j, \forall j \in \mathcal{M}
\end{equation}

As a consequence of Theorem 1, the energy consumption decreases if $q_{i,j}$ is increased. Thus, for the optimal solution, equality is achieved in \eqref{modified C2_new_q}, which means substituting (\ref{Pq}b) with \eqref{modified C2_new_q} will not change the solution.
Then, for inequality constraint \eqref{modified C2_new_q}, its second derivative is $\sum_{i \in \mathcal{K} } \frac{2 \eta L_{i,j}}{(D_i -t_{i,j})}>0$, and thus, it is convex. For the objective function, we have $$\frac{\partial^2 E_{i,j}}{\partial t_{i,j}^2}=  \frac{N_0 L_{i,j}^2}{h_{i,j}x_{i,j} t_{i,j}^3} 2^{\frac{L_{i,j}}{x_{i,j} t_{i,j}} }  \ln2^2>0  .$$ Therefore, \eqref{Pq} with the relaxed constraint \eqref{modified C2_new_q} is convex.
\end{enumerate}
\end{IEEEproof}

Now let us consider the joint optimization problems, where one of the variables is fixed, while the other two are optimized jointly.

\begin{theorem}\label{theorem-L-opt}
Subproblems of P2, where $\mbf{L}$ is one of the decision variables, are non-convex problems.
\end{theorem}
\begin{IEEEproof}
\begin{enumerate}[label=\alph*)]
\item Optimizing data size $\mbf{L}$ and transmission bandwidth $\mbf{x}$ under fixed processing power allocation $\mbf{q}$. {\color{black} This has been shown in Theorem \ref{theorem_main_conv}, namely, the proof of the non-convexity of P2.}
\item Optimizing data size $\mbf{L}$ and processing power $\mbf{q}$  under fixed  transmission bandwidth allocation $\mbf{x}$. In this case, the Hessian matrix is given by
\[
\mbf{\vec{H}}_{i,j} = \frac{ N_0  x_{i,j}}{h_{i,j}}
   \begin{bmatrix}
    \mbf{\vec{H}}_{i,j}(1,1)
     & \mbf{\vec{H}}_{i,j}(1,2) \\
    \mbf{\vec{H}}_{i,j}(2,1)
     & \mbf{\vec{H}}_{i,j}(2,2)
  \end{bmatrix},
\]
where $ \mbf{\vec{H}}_{i,j}(1,1) = \frac{ D_{i}^2}{x_{i,j}^2(D_{i}-\eta L_{i,j} /q_{i,j})^3} 2 ^{\frac{L_{i,j}/x_{i,j}}{D_{i}-\eta L_{i,j} /q_{i,j}}} \ln2^2 $, $\mbf{\vec{H}}_{i,j}(1,2) =\mbf{\vec{H}}_{i,j}(2,1)=  ( \frac{\eta }{q_{i,j}^2 } + \frac{ \eta^2 L_{i,j}^2 \ln2/x_{i,j} - \eta D_i L_{i,j} q_{i,j} \ln2/x_{i,j} }{q_{i,j} (D_i q_{i,j} - \eta L_{i,j} )^2 } - \frac{\eta D_i L_{i,j}^2 q_{i,j} \ln2^2/x_{i,j}^2  }{(D_i q_{i,j}- \eta L_{i,j} )^3 } ) 2 ^{\frac{L_{i,j}/x_{i,j}}{D_{i}-\eta L_{i,j} /q_{i,j}}} - \frac{\eta}{q_{i,j}^2} $ and $\mbf{\vec{H}}_{i,j}(2,2) =  ( -\frac{2 \eta L_{i,j} }{q_{i,j}^3 } + \frac{ 2\eta D_i L_{i,j}^2 \ln2/x_{i,j}  }{q_{i,j} (D_i q_{i,j} - \eta L_{i,j} )^2 } - \frac{\eta^2  L_{i,j}^4  \ln2^2/x_{i,j}^2  }{q_{i,j}(D_i q_{i,j}- \eta L_{i,j} )^3 } ) 2 ^{\frac{L_{i,j}/x_{i,j}}{D_{i}-\eta L_{i,j} /q_{i,j}}} + \frac{2 \eta L_{i,j}}{q_{i,j}^3} $.

It can be verified that $\rm{det}(\mbf{\vec{H}}_{i,j}) \geq 0$ does not always hold, which indicates the problem is non-convex. For example, by setting $ D_i=1, \eta=1, L_{i,j}=2, x_{i,j}=1, q_{i,j}=2.1, \frac{N_0}{h_i}=1 $, we obtain $\rm{det}(\mbf{\vec{H}}_{i,j})=-28.7<0 $.
\end{enumerate}
\end{IEEEproof}

However, the third subproblem of two decision variables, that is, the case when the size of the data blocks is fixed, is more tractable, according to the following theorem.

\begin{theorem}\label{theorem-L-fixed}
The subproblem of P2, where $\mbf{L}$ is fixed and the transmission bandwidth $\mbf{x}$ and processing power $\mbf{q}$ allocation needs to be optimized, is a convex problem.
\end{theorem}
\begin{IEEEproof}
We replace variables $q_{i,j}$ with $t_{i,j}=D_i-{\eta L_{i,j}}/{q_{i,j}} $ to get
\begin{subequations} \label{Pq_2}
\begin{align}
 \underset{ {\mbf{t}}}{\text{min}} & ~ \sum_{i \in \mathcal{K}} \frac{  N_0 }{h_{i,j}} x_{i,j} t_{i,j} \left(2^{\frac{L_{i,j}}{x_{i,j} t_{i,j}}}-1 \right) \\
\text{s.t.}
& ~\sum_{i \in \mathcal{K}}  x_{i} =B  \\
& ~ \sum_{i \in \mathcal{K} } \frac{\eta L_{i,j}}{D_i-t_{i,j}} = C_j, \forall j \in \mathcal{M}
\end{align}
\end{subequations}

First, equality constraint (\ref{Pq_2}b) is affine. (\ref{Pq_2}c) can be relaxed to \eqref{modified C2_new_q}, which is convex. Last, let us consider the objective function (\ref{Pq_2}a). The Hessian matrix is given by
\[
\mbf{\hat{H}}_{i,j} =
  \frac{N_0}{h_{i,j}} \cdot \begin{bmatrix}
    \mbf{\hat{H}}_{i,j} (1,1)
     & \mbf{\hat{H}}_{i,j} (1,2) \\
    \mbf{\hat{H}}_{i,j} (2,1)
     & \mbf{\hat{H}}_{i,j} (2,2)
  \end{bmatrix},
\]
where $\mbf{\hat{H}}_{i,j} (1,1)=\ln2^2 \cdot 2^{\frac{L_{i,j}}{ t_{i,j} x_{i,j}} } \cdot \frac{L_{i,j}^2}{t_{i,j} x_{i,j}^3}$, while $\mbf{\hat{H}}_{i,j} (2,2)=\ln2^2 \cdot 2^{\frac{L_{i,j}}{x_{i,j} t_{i,j}} } \cdot \frac{L_{i,j}^2}{x_{i,j} t_{i,j}^3}$. Besides, $\mbf{\hat{H}}_{i,j} (1,2)=\mbf{\hat{H}}_{i,j} (2,1)=2^{\frac{L_{i,j}}{x_{i,j} t_{i,j}} }-1 + \ln2^2 \cdot \frac{L_{i,j}^2}{x_{i,j}^2t_{i,j}^2} 2^{\frac{L_{i,j}}{x_{i,j} t_{i,j}} }- \ln 2 \cdot \frac{L_{i,j}}{x_{i,j} t_{i,j}} 2^{\frac{L_{i,j}}{x_{i,j} t_{i,j}} }$.
After some algebraic manipulations, it can be verified that $\rm{det}(\mbf{\hat{H}}_{i,j})>0$ holds for all $\frac{L_{i,j}}{x_{i,j} t_{i,j}}>0$, which indicates (\ref{Pq_2}a) is convex. This completes the proof.
\end{IEEEproof}

\begin{corollary}\label{col-single-fixed-AP}
Consider the case, when each node can connect to a single AP only, and this AP is predefined (e.g., the one with best SNR). Then, the problem of transmission bandwidth and processing resource allocation is convex.
\end{corollary}
\begin{IEEEproof}
This is a special case addressed by Theorem \ref{theorem-L-fixed}, where $\mbf{L}_i$ has only one non-zero element  $\forall i \in \mathcal{K}$.
\end{IEEEproof}

\subsection{Iterative Resource Allocation for Multi-AP Processing}
\begin{algorithm}
\caption{Iterative Resource Allocation for Multi-AP Processing}
\label{alg:iterative}
\begin{algorithmic}[1]
\State {\textbf{Initialization:}}
$L_{i,j} \leftarrow L_{i,j}^{(0)}, \forall i \in \mathcal{K}, j \in \mathcal{M}$;
\State Update  $x_{i,j}, q_{i,j}, \forall i \in \mathcal{K}, j \in \mathcal{M}$ based on BCAA, and calculate $\sum_{i=1}^K \sum_{j=1}^M  E_{i,j}^0$;
\State $\sum_{i=1}^K \sum_{j=1}^M  E_{i,j}^1 \leftarrow \sum_{i=1}^K \sum_{j=1}^M  E_{i,j}^0+ 2 \epsilon$;
\State \textbf{while} $\sum_{i=1}^K \sum_{j=1}^M  E_{i,j}^1 - \sum_{i=1}^K \sum_{j=1}^M  E_{i,j}^0 > \epsilon$ \textbf{do}
\State \hspace{20pt} Update $L_{i,j}$ based on DAA, and recalculate $\sum_{i=1}^K \sum_{j=1}^M  E_{i,j}^1$;
\State \hspace{20pt} Update $x_{i,j}, q_{i,j}$ based on BCAA, and recalculate $\sum_{i=1}^K \sum_{j=1}^M  E_{i,j}^0$;
\State \textbf{end while}
\end{algorithmic}
\end{algorithm}

In this section, an Iterative Resource Allocation algorithm is proposed to solve the non-convex joint resource allocation problem P2.
As shown in \text{Algorithm \ref{alg:iterative}}, the proposed algorithm follows two iterative steps: i) the Data Allocation Algorithm (DAA) updates $\mbf{L}$ to allocate the data, for given bandwidth and computing resource allocation $\mbf{x}$ and $\mbf{q}$, and ii) the Bandwidth and Computing resource Allocation Algorithm (BCAA) updates $\mbf{x}$ and $\mbf{q}$ to allocate bandwidth and computing resource, for given $\mbf{L}$.
We denote by $E_i^{t}$ and $E_i^{x}$ the energy consumption of user $i$ after optimizing $t_i$ and $x_i$, respectively, and $\epsilon$ is the stop condition.
Additionally, $L_{i,j}^{(0)}$ denotes the initial value for $L_{i,j}$, and it can be obtained using a fixed allocation, e.g., equal allocation, or a random allocation, e.g., following a uniform distribution.

{\bf{The Data Allocation Algorithm (DAA):}}
The data allocation problem under given bandwidth and computing resource is shown to be convex in {\color{black} Theorem~\ref{theorem-singleparam}.}
Therefore, we can use the Karush-Kuhn-Tucker (KKT) condition to derive the
optimal $L_{i,j}$. The KKT condition for data $L_{i,j}$ is given by \eqref{lambda} at the top of next page.
\begin{figure*}[!t]
\normalsize
\begin{equation} \label{lambda}
g( L_{i,j})=\frac{N_0 x_{i,j} }{ h_{i,j} }  \left[ \left( \frac{D_{i} \ln2}{x_{i,j}(D_{i}-\eta L_{i,j} /q_{i,j})} - \frac{\eta}{q_{i,j}} \right) 2 ^{\frac{L_{i,j}/x_{i,j}}{D_{i}-\eta L_{i,j} /q_{i,j}}} + \frac{\eta}{q_{i,j}}  \right]+ \lambda=0.
\end{equation}
\end{figure*}
Note that in \eqref{lambda} $\lambda$ is the Lagrange dual variable.

For given $\lambda$, the above equation can be used to obtain $L_{i,j}$. Specifically, we have $\frac{\partial g( L_{i,j})}{\partial L_{i,j} }= \frac{N_0 D_{i}^2}{h_{i,j}x_{i,j}(D_{i}-\eta L_{i,j} /q_{i,j})^3} 2 ^{\frac{L_{i,j}/x_{i,j}}{D_{i}-\eta L_{i,j} /q_{i,j}}} \ln2^2 >0.$, which indicates that $g( L_{i,j})$ grows with $L_{i,j}$, and thus a bisection search can be used to obtain $L_{i,j}$ by comparing $g(L_{i,j})$ with 0.
Now the problem lies in how to obtain $\lambda$. When $\lambda$ is increased, $L_{i,j}, \forall j \in \mathcal{M}$ will decrease to ensure $g(L_{i,j})=0$. Meanwhile, $\sum_{j=1}^M L_{i,j}= L_i$ needs to hold. Consequently, $\lambda$ can also be obtained with bisection search, by comparing $\sum_{j=1}^M L_{i,j}$ with $L_{i}$.

The resulting DAA consists of two loops: an outer loop to find the value of $\lambda$ and an inner loop to determine the data allocation ${\bf{L}}_{i}$.
The computational complexity is $O \left(  K M  \log_2(\lambda ) \log_2(L_i)   \right)$

{\bf{The Bandwidth and Computing resource Allocation Algorithm (BCAA):}}
{\color{black}
Under given data allocation, the joint bandwidth and computing resource allocation problem is shown to be a convex one in Theorem~\ref{theorem-L-fixed}, and thus, the optimal solution can be obtained using convex optimization tools, e.g., interior-point method. Nonetheless, considering that the bandwidth allocation is global {\color{black}for the entire network}, while the computing resource allocation is local at each AP, we propose {\color{black}to apply} an iterative algorithm to solve the joint bandwidth and computing resource allocation.
{\color{black}The algorithm is presented in detail in our previous work \cite{Zeng_WCL2019}, here we summarize it briefly.}
The proposed algorithm consists of two iterative steps: i) the Bandwidth Allocation Algorithm (BAA) updates $\mbf{x}$ to allocate bandwidth across and within the APs, for given $\mbf{t}$ ($t_{i,j}=D_i-{\eta L_{i,j}}/{q_{i,j}} $), and ii) the Computation resource Allocation Algorithm (CAA) updates $\mbf{t}$ to allocate the computing resource at each AP, for given bandwidth allocation $\mbf{x}$.

{\bf{Bandwidth Allocation Algorithm (BAA):}}
Assume that the computing resource allocation $\mbf{t}$ is given. Then the KKT condition for $x_{i,j}$ is given by
\begin{align} \label{beta}
f(x_{i,j}) &= \frac{ N_0 t_{i,j}}{h_{i,j}} \left[ 2^{\frac{L_{i,j}}{t_{i,j} x_{i,j}}} -    \frac{ L_{i,j}}{t_{i,j} x_{i,j}}  2^{\frac{L_{i,j}}{t_{i,j} x_{i,j}} } \ln2-1 \right] + \beta \\ \nonumber
&=0,
\end{align}
where $\beta$ is the introduced auxiliary variable, satisfying $\beta>0$. As for $L_{i,j}$ in \eqref{lambda}, the bisection method can be used to obtain $\beta$ and $x_{i,j}$. Likewise, the resulting BAA consists of two loops: an outer loop to find the value of $\beta$ and an inner loop to determine the bandwidth allocation $\bf{x}$.
The computational complexity is $O( KM \log_2(B) \log_2(\beta ))$.

{\bf{Computing resource Allocation Algorithm (CAA):}}
Under given bandwidth allocation, the computing resource allocation is independent across the APs. Thus, the energy minimization for each AP is equivalent to that of the overall system.
Let us consider  AP $j$, $j\in \mathcal{M}$. The KKT condition for $t_{i,j}$ is given by
\begin{align*}
 v(t_{i,j})& = \frac{ N_0 x_{i,j}}{ h_{i,j}} \left[ 2^{\frac{L_{i,j}}{x_{i,j} t_{i,j}}} -   \frac{L_{i,j}}{x_{i,j} t_{i,j}} 2^{\frac{L_{i,j}}{x_{i,j} t_{i,j}}} \ln2 -1 \right] \\
 &~~+ \frac{W_{i,j} }{(D_{i}-t_{i,j})^2} \mu_j \\
&=0,
\end{align*}
where $\mu_j$ is the introduced auxiliary variable, satisfying $\mu_j \geq 0$.
As for $L_{i,j}$ and $x_{i,j}$, the bisection method can be used to obtain $\mu_j$ and $t_{i,j}$. Likewise, the resulting CAA consists of two loops: an outer loop to find the value of $\mu_j$ and an inner loop to determine the computing resource allocation $t_{i,j}$.
The computational complexity is $O( KM \log_2(D_i) \log_2(\mu_j ))$.

{\color{black}BAA and CAA have to be repeated until convergence}. Since the bandwidth and computing resource allocation problem is convex, convergence is guaranteed and the converged local optimum is also the global optimum.

}


{\bf{Convergence and Complexity:}}
\begin{theorem}
The Iterative Resource Allocation algorithm converges in finite steps.
\end{theorem}

\begin{IEEEproof}
In both lines 9 and 10 of Algorithm \ref{alg:iterative}, the energy consumption decreases, or remains unchanged. Since there is a lower bound for the energy consumption, e.g., 0, the Iterative Resource Allocation algorithm always terminates, either by reaching the lower bound, or by achieving a decrease less than $\epsilon$. Therefore, convergence is guaranteed.
\end{IEEEproof}

Denote the number of iterations required for Algorithm \ref{alg:iterative} {\color{black}and BCAA  to converge by $I$ and $I_0$, respectively. The total computational complexity is $O( I KM [ \log_2(L_i) \log_2(\lambda ) + I_0 \log_2(B) \log_2(\beta)+ I_0\log_2(D_i) \log_2(\mu_j) ] $.

\begin{figure}
\centering
\includegraphics[width=1\linewidth]{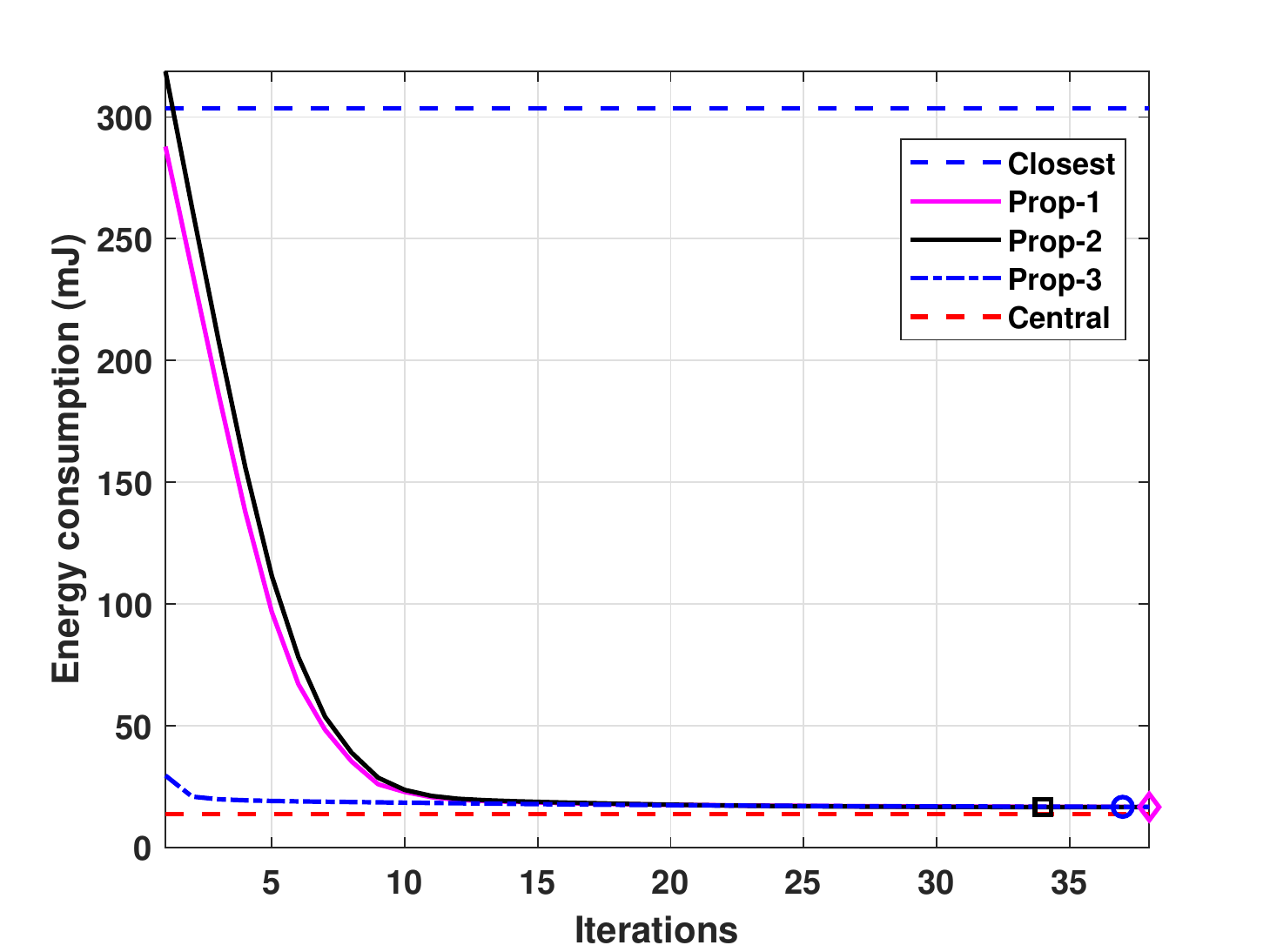}
\caption{Convergence of the Iterative Resource Allocation algorithm under different network scenarios, and different initial user data allocation. {\color{black}The markers show the termination of the iteration for the corresponding algorithms.}}\label{iteration}
\end{figure}

\section{Numerical Results}

In this section numerical results are presented to evaluate the effectiveness of the proposed energy minimization algorithm. Specifically, first we evaluate the convergence of the Iterative Resource Allocation algorithm. Then we evaluate what is the level of distributed processing under the optimized resource allocation, to see whether this possibility leads to significant performance gains.

We consider a  small cell multi-AP multi-user scenario, where image processing tasks are offloaded to the edge computing servers.  The network parameters are set as for example in \cite{3gpp_pl, Cyou, Ymao}, while the
task requirements follow \cite{5G_Ini, Cyou, A_kiani}.
More specifically, the number of APs and users are set to $M=4$ and $K=8$. 
The users are placed uniformly randomly within the entire coverage region.
The pathloss model follows $30.6+36.7\log_{10}(d)$, where $d$ is the distance in meter. The total bandwidth is 10 MHz, while the thermal noise density is -174 dBm/Hz.
The data size of the computing task is 1.5 Mbits, while the delay constraint is 0.5 s. The computing coefficient $\eta$ is $10^3$, and thus the computing need is 1.5 G CPU cycles. The computing capacity at each AP is 25 G CPU cycles/s.


Let us first investigate how the proposed iterative algorithm, described in Algorithm 1, converges. We consider three different initializations of the user data: {\emph{Prop-1} denotes the case when the data for each user are equally allocated among the APs; \emph{Prop-2} is obtained following a uniformly random data size allocation; and finally \emph{Prop-3} represents the case when $90\%$ of the data are allocated to the AP with the best channel condition, while the rest are equally allocated to the remaining APs. The stop condition is $\epsilon=10^{-2}$ mJ.
According to the results, the algorithm converges to the same solution, independently from the initial data allocation.
This indicates that the proposed algorithm is robust to the initialization. Moreover, \emph{Prop-3} {\color{black}requires the least number of iterations to approach the converged solution in all scenarios}, which implies the proposed algorithm converges to a solution where users associate most of the data to their best APs.

\begin{figure}
\centering
\includegraphics[width=1\linewidth]{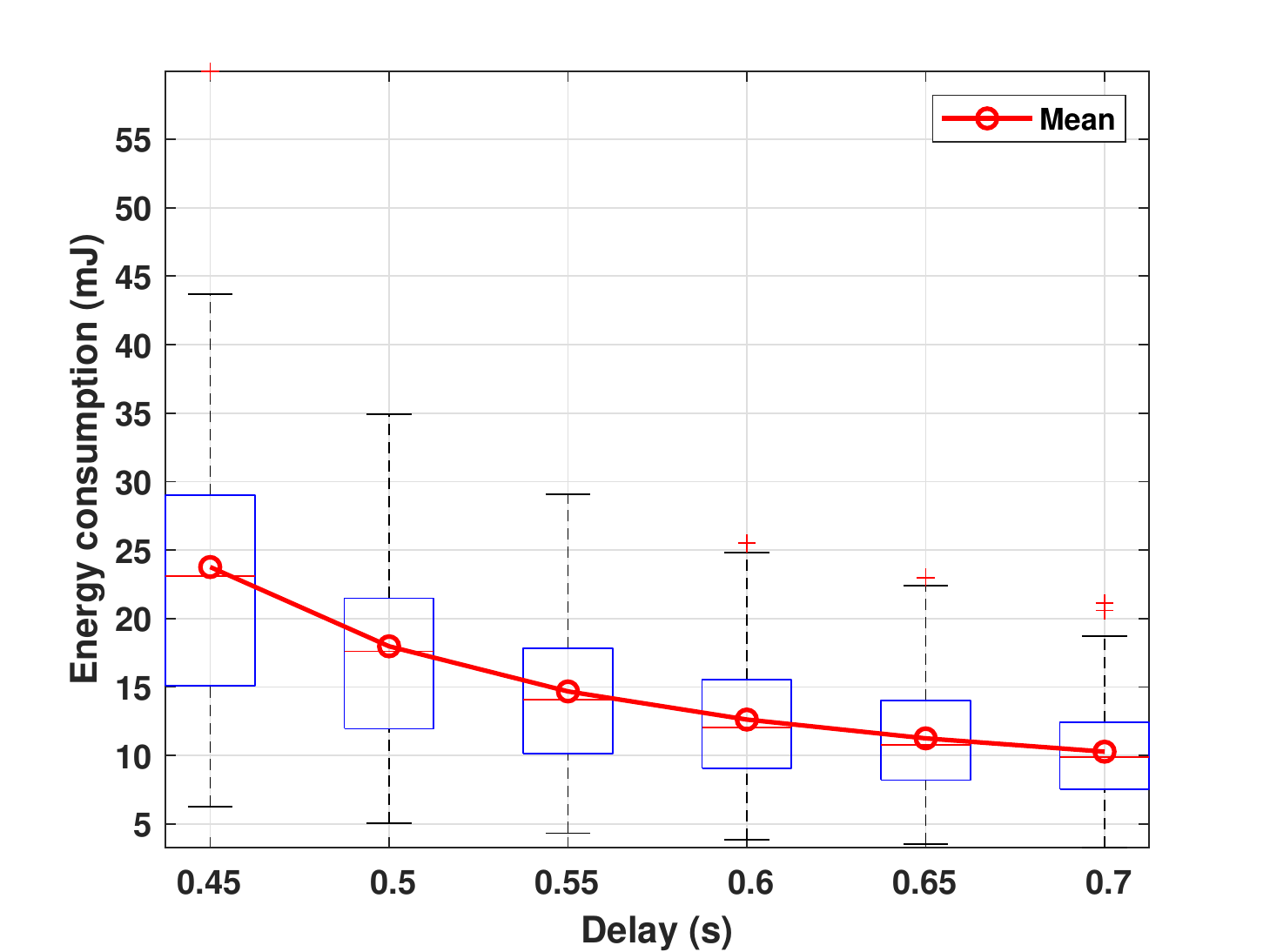}
\caption{Energy consumption versus delay under different network scenarios.}
\label{Delay}
\end{figure}

We then plot the energy consumption versus the delay requirement in Fig.~\ref{Delay}. As expected, the energy consumption decreases as the delay requirement becomes looser, for all considered scenarios.  
Lastly, we investigate how much the opportunity of parallel processing is utilized in the allocations. 
Fig.~\ref{Load} considers only the users that connect to more than one AP, and shows what is the largest share of load sent to one of them, {\color{black}namely $\underset{ j} {\max}~ L_{i,j}/L_i $}. The values are rather high for all cases, showing that most of the users have a preferred AP. This validates why an unbalanced initialization as \emph{Prop-3} on Fig. \ref{iteration} leads to fast convergence for the Iterative Resource Allocation.

\begin{figure}
\centering
\includegraphics[width=1\linewidth]{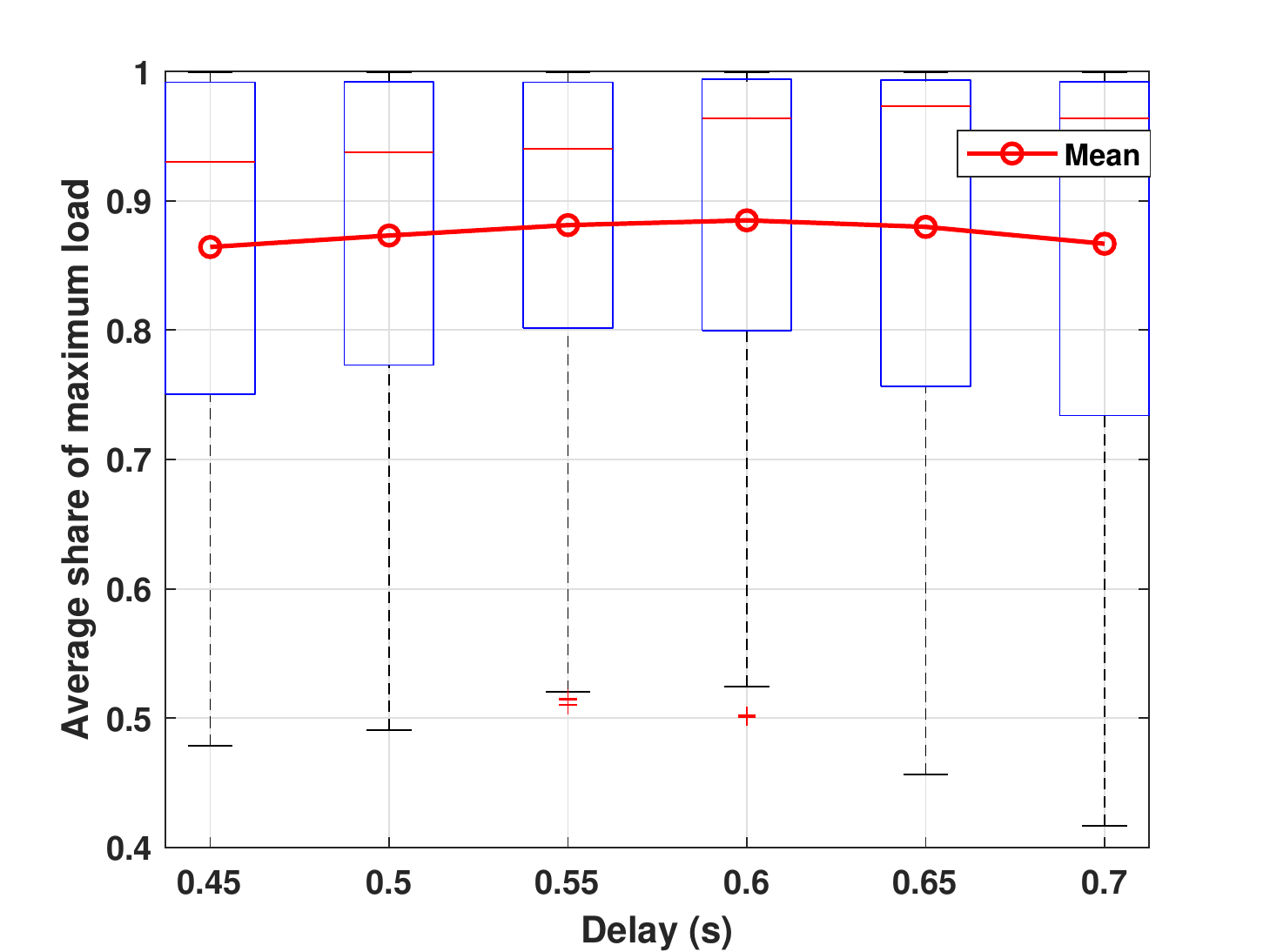}
\caption{Ratio of largest load share versus delay under different network scenarios.}
\label{Load}
\end{figure}

\section{Conclusion}
In this paper we investigated the join wireless and computing resource allocation for a multi-AP multi-user MEC network. The general case with parallel processing and global bandwidth sharing was considered, and the system objective was to minimize the sum transmission energy under response time constraints. The formulated problem was shown to be non-convex. To address it, we first investigated the complexity of optimizing a part of the system parameters, and based on the results proposed an Iterative Resource Allocation procedure with guaranteed convergence. Presented numerical results show that the proposed iterative algorithm converges rapidly to the local optimum. Moreover, by comparing the proposed iterative algorithm with the lower and upper bounds, it is clear that free selection of APs is crucial for obtaining decent system performance.

\balance

\bibliographystyle{IEEEtran}
\bibliography{IEEEabrv,conf_short,jour_short,mybibfile}
\end{document}